\documentclass[a4,11pt]{article}
\usepackage{harvard}
\usepackage{graphicx}

\begin{document}
\bibliographystyle{kluwer}

\pagestyle{myheadings}
\markright{Nicholson et. al.}

\title{Modelling Horizontal and Vertical Concentration Profiles of
Ozone and Oxides of Nitrogen within High-Latitude Urban Areas}

\author{
 James P. Nicholson and Keith J. Weston\\Department of Meteorology, University
of Edinburgh, Edinburgh, EH9 3JZ, U.K.\\
\and
  David Fowler\\Centre for Ecology and Hydrology, Bush Estate,
Midlothian, EH26 0QB, U.K.}

\maketitle

\pagenumbering{arabic}


\renewcommand{\baselinestretch}{1.0}
\small\normalsize

\begin{abstract}
Urban ozone concentrations are determined by the balance between ozone
destruction, chemical production and supply through advection and turbulent
down-mixing from higher levels. At high
latitudes, low levels of solar insolation and high horizontal
advection speeds reduce photochemical production
and the spatial ozone concentration patterns are
largely determined by the reaction of ozone with nitric oxide and dry
deposition to the surface.
A Lagrangian column model has been developed to simulate the mean
(monthly and annual) three-dimensional structure in ozone and 
nitrogen oxides ($NO_{x}$) concentrations in the boundary-layer 
within and immediately around an urban areas.
The short time-scale photochemical processes  of ozone and $NO_{x}$, as well
as emissions and deposition to the ground, are simulated.
 
The model has a horizontal resolution of 1x1km and high resolution in
the vertical. It has been applied over a 100x100km
domain containing the city of
Edinburgh (at latitude $56^\circ$N) to simulate the city-scale
processes  
of pollutants. 
Results are presented, using averaged wind-flow
 frequencies and appropriate stability conditions, to show the extent of
the depletion of ozone by city emmisions.
The long-term average spatial patterns in the surface ozone and $NOx$
concentrations over the model domain are reproduced quantitatively. 
The model shows the average surface ozone concentrations in the urban
area to be lower than the surrounding rural areas by typically $50\%$
and that the areas experiencing a $20\%$ ozone depletion are
generally 
restricted to within the urban area. The depletion of the ozone
concentration to less than $50\%$ of the rural surface values extends only 20m
vertically above the urban area.
A series of monitoring sites for ozone, nitric oxide and nitrogen dioxide on a
north-south transect through the city - from an urban, through a
semi-rural, to a remote rural location - allows the comparison of
modelled
with observed data for the mean diurnal cycle of ozone concentrations. In the
city-centre, the cycle is well reproduced, but  the ozone concentration is
consistently overestimated. 
\end{abstract}

{\em Key-word index:} Tropospheric ozone, Lagrangian column model, urban,
nitrogen oxides, vertical exchange.

\section{Introduction}
Tropospheric ozone is a photochemical oxidant formed largely by photochemical
reactions. 
In the troposphere ozone acts as a greenhouse gas \cite{Fishman}, 
\cite{Chalita} and is toxic to plants, reducing crop yields
\cite{Hewitt}, and to humans as a respiratory irritant \cite{WHO},
as well as damaging
both natural and man-made materials, such as stone, brick-work and rubber,
\cite{PORG93}.
Quantifying the dose and exposure of the human population, vegetation
and materials to ozone is required to assess the scale of ozone impacts
and to develop control strategies.

A network of 17 rural ozone monitoring stations across the UK
provides broad-scale regional spatial patterns in tropospheric ozone
concentrations in rural areas \cite{PORG93}. Peak ozone concentrations increase
from north to south across the UK as the south has higher concentrations of
the primary pollutants necessary for ozone production (from greater emissions
and proximity to continental sources) as well as greater
frequency of meteorological conditions suitable for ozone production. Mean ozone concentrations
also increase with altitude \cite{PORG97} and are higher in a 5-10km coastal strip \cite{Entwistle}.
The mean annual-average background concentration of ozone for the UK is
approximately 
50$\mu$g m$^{-3}$, though there is a wide variation for individual 
episodes about this value, ranging up to about 400$\mu$g m$^{-3}$ \cite{PORG93}. 
There are clear diurnal and annual cycles in ozone concentrations in the
UK, with a mid-afternoon peak and nocturnal minimum and a spring maximum
and autumn minimum. The diurnal cycle illustrates the fundamental
importance of vertical mixing \cite{Garland}.

Ozone concentrations in urban areas are of particular interest and importance
as the population is largely urban-based and ozone concentrations show
greater spatial and temporal variations in urban areas. Ozone
concentrations are smaller in urban areas than they are in surrounding
rural areas due to the reaction of ozone with nitric oxide, emitted from
combustion sources, forming nitrogen dioxide. Where air is allowed to
stagnate over an urban area, the effects of strong insolation and
accumulating ozone precursors can produce very high ozone concentrations.
An example of this is the photochemical smog that affects the
Los Angeles area of California, where a combination of meteorology, local topography
and very high pollutant emission levels produce
dangerously high ozone concentrations on many days of the year 
\cite{Lents}. However, in much of the UK and other high-latitude areas where
insolation levels are lower and wind-speeds are larger (maintaining a
steady advection of air through the urban air-shed), there is
a different spatial distribution of concentrations with 
annual mean ozone concentrations generally smaller in urban areas.

There are many urban sites at which ozone is monitored in the UK \cite{PORG97}.
Attempts to map ozone
concentrations using both rural and urban
monitoring are complicated by the interaction of local chemistry with the
larger-scale meteorological factors determining ozone concentrations. 
Recent studies on higher resolution ozone mapping have used an urbanisation index 
\cite{PORG97}, but these have not been able to account for the movement
of ozone and ozone-depleted air into and out of the urban areas.

Various studies have examined the characteristics of ozone and
nitrogen oxide concentrations around urban areas 
\cite{Ball}. A large number of studies have focussed on the
two-dimensional structure of the `urban plume' of photochemical-ozone downwind
of a city, with different cities around the world being studied
\cite{Cleveland}, \cite{White}, \cite{Varey}, \cite{Lin} and
\cite{Silibello}. There have been few studies
of the three-dimensional mean structure of ozone concentrations in urban areas,
especially at high latitudes, where ozone production is
of secondary importance in
describing the spatial distributions around cities. These studies are largely
interpretations of observational
data \cite{Angle} and \cite{Leahey}.

A boundary-layer Lagrangian column model has been developed to simulate
the mean three-dimen\-sional structure
in ozone and nitrogen oxide concentrations in the boundary-layer
within and immediately around high-latitude urban areas at a spatial scale of
1x1km. (The model is not appropriate for use with ``real time''
trajectories.) The model 
simulates the effects of ozone depletion at the surface as a consequence
of the reaction with emitted nitric oxide and dry deposition to the 
surface.
This has been used to follow a range of one-dimensional trajectories
over a distance of $10^{5}$m and a travel time in the order of $10^{4}$s,
through a simulated city under a variety of meteorological and pollutant
emission regimes representing seasonal and diurnal extremes. An assessment
of the extent of ozone destruction occurring,
the rate of recovery of surface ozone concentrations downwind of the city and
the influence of meteorological parameters on the ozone
concentration has been provided using the model.
 
The model has been applied over a 100x100km domain containing
a simulation of the emission field over the city of
Edinburgh. Edinburgh was used as a generic, high-latitude
city for modelling purposes.  A land-use array has been created as input to the model
with spatially- and temporally-variable emission and deposition values.

\section{Ozone Destruction by Nitric Oxide}

High latitude cities are, in general, well-ventillated, so that the
timescale for air traversing the city is small compared to that for
ozone generation. Under these conditions, the three reactions that 
are fundamental in the determination of ozone ($O_{3}$) concentrations 
in urban areas \cite{Wayne} are: 
 
\begin{equation} O + O_{2} + M  \rightarrow  O_{3} + M	\hspace{1.8cm}  k_{1}
\label{O+O2}
\end{equation}

\begin{equation} NO_{2} + \it{h\nu}  \rightarrow  NO + O  \hspace{1.8cm} J_{2}
\label{NO2phot}
\end{equation}

\begin{equation} NO + O_{3}  \rightarrow  NO_{2} + O_{2}  \hspace{1.8cm} k_{3}
\label{NO+O3} 
\end{equation}

where $k_{1}$ and $k_{3}$ are reaction rate constants and $J_{2}$
is the photolysis rate for nitrogen dioxide ($NO_{2}$).

In the UK, the main source of nitrogen oxides is the combustion of
fossil fuels, $\it{ie.}$ the burning of coal, oil and gas   
in power stations and the combustion of petrol and diesel by road traffic \cite{Salway}.
The largest source in large urban areas is the exhaust from road vehicles.
The emissions are mainly $NO$ with ratios of $NO$:$NO_{2}$
in excess of about 3:1 \cite{Selles}.
A large amount of this $NO$ is rapidly oxidised by ozone (Reaction \ref{NO+O3}).

The diurnal cycle of urban ozone concentrations show a
maximum concentration at night from 0100 to 0500 hours when traffic density, and hence
$NO$ emissions, are lowest \cite{PORG97}. Minimum concentrations occur when
the $NO$ concentration peaks during the morning and evening rush hours.
Wind-speed also has an effect on urban ozone concentrations with
ozone-rich air advected into the city from the
surrounding countryside in windy condition. At the same time, $NO$ emitted
in the city is diluted in the well-mixed air and so Reaction (\ref{NO+O3}) is
less dominant - reducing both $O_{3}$ destruction and $NO_{2}$
production \cite{Oke}.

\section{Experimental Method}

\subsection{Description of the Model Used}
The TERN model (Transport over Europe of Reduced Nitrogen) was developed
in the early 1990s \cite{ApSimon} to examine the release, transport and
chemistry of ammonia. It is a Lagrangian
column model with detailed vertical resolution and diurnally varying
emissions, deposition and turbulent mixing. It was further developed to
examine ammonia and ammonium transport and deposition over the UK \cite{Singles}. It can
be used to obtain
vertical concentration profiles of pollutants emitted or generated
at or near the surface and transported over long distances on a single
trajectory.

This Lagrangian column model was considered suitable as a basis for a model to
be used in this study of urban ozone concentrations because it allows detailed vertical mixing
to be considered and thus is able to represent the marked variations with height of the
concentration profiles. An Eulerian model with a comparable run-time would not
give the required vertical resolution \cite{ApSimon}.
The TERN model has been considerably modified so as to be more suitable
to modelling ozone concentrations in a city but the basic processes of
calculating the extent of vertical mixing and the exchange between the vertical layers have remained
the same.
The detailed vertical resolution was used to provide a
sophisticated treatment of the vertical diffusion of the pollutant species.
The column of air that is advected over the surface (in the absence of
vertical shear in the column) is the lowest 2.5km of the
troposphere and is divided into 33 layers. There is much
finer resolution near the surface (the lowest five layers are of depth
1, 2, 2, 4 and 15m), allowing more detail where the highest
concentration gradients occur. The neglect of lateral dispersion, both
by shear and turbulence, is not a serious omission for two reasons:
firstly, the model is principally used to determine mean fields, for
which climatological wind data are used, covering the whole 360 degrees;
and secondly, the relatively small spatial scale of both the application
of the model (and the resulting fields), means that changes due to
lateral mixing are small compared to those due to vertical mixing.

Parameterisation of the boundary layer includes the effects on
the mixing layer depth and the vertical diffusivity of the
meteorological variables - insolation, cloud cover and wind-speed.
The mixing layer depth is calculated on a diurnally varying pattern
depending on the time of day and the prevailing meteorological
conditions. Carson's model \cite{Carson} is used to 
calculate the development during the day, with a mechanical mixing factor
proportional to the geostrophic wind-speed. At night, the 
depth is determined from a combination of
the Pasquill category and the wind-speed \cite{Pasquill}.

The vertical mixing of gases, is determined using the diffusion equation:

\begin{equation} 
\frac{\delta C}{\delta t} = \frac{\delta}{\delta z} \left(K_{z}\frac{\delta C}{\delta z}\right)
\end{equation}

The coefficient of mixing, $K_{z}$, is defined as a function of height for different stability
conditions. In the model, $K_{z}$ is assumed to linearly increase with
height to a value of $K_{max}$ at height $z_{m}$ and to be constant
above this height up to the top of the mixed layer. During the day,
$z_{m}$ is fixed at 200m and $K_{max}$ is dependent on the larger of two
terms, one representing mechanical mixing and the second on convective
mixing. The first of these terms is dependent on stability and wind
speed and the second on surface heat flux.  At night, both $K_{max}$ and
$z_{m}$ are dependent on wind speed and cloud cover \cite{ApSimon}.

TERN's treatment of mechanical turbulence takes no account of the surface
roughness. However, in
the city the increased roughness effect of the buildings would increase
the turbulence of the air compared to the countryside. 
Assuming a logarithmic
wind profile over the lowest 200m (with no zero-plane displacement) and surface roughnesses of 0.05m for
rural and 1.0m for urban areas \cite{Stull}, the ratio of urban to rural friction
velocities was calculated to be 1.5. This factor is used as a scaling
factor for $K_{z}$ to simulate enhanced mixing over urban areas.

A diurnally-varying dry deposition velocity ($V_{D}$), is included for 
nitrogen dioxide and ozone \cite{Hargreaves}.

The chemistry employed in the model is very simple, with
ozone destruction and production by Reactions (\ref{O+O2}-\ref{NO+O3})
included but ozone production through hydrocarbon degradation ignored. 
Photodissociation (Reaction 2) is dependent on radiation levels which
are parameterised in terms of time of day, time of year and
climatological cloud cover.

The time-step used can be varied over a wide range of values but the
chemistry involved requires time-steps of a few seconds to be 
treated accurately. The time-step used for all model runs was
1.5 seconds (or 40min$^{-1}$). The chemistry would actually have allowed
for a longer time-step than this, which would have been desirable in
keeping the run-time down, but the differential equations
used to calculate vertical diffusion became unstable at low
wind-speeds with longer time-steps.

\subsection{Edinburgh Data}

A large part of this study has been performed on a model of the city of
Edinburgh: a city of 450,000 inhabitants on the east coast of Great
Britain at a latitude of $56^{\circ}$. The city is set on a coastal     
plain between the Firth of Forth to the north and a range of 600m hills
to the south. It is a centre of finance and service industries and as
such does not have a large number of factories in the city centre or nearby. It 
does however
have a rapidly expanding road vehicle fleet that generates $NO_{x}$. 
It is also a compact city, approximately 10km in diameter, with little  
suburban sprawl or outlying towns. The nearest large urban area is
Glasgow some 70km to the west. The omission of these distant sources is
discussed in section 7.

\subsection{Nitrogen Oxide Emissions}

The $NO_{x}$ emissions
in the model comprises 1x1km data of Edinburgh emissions for a 12x10km
grid in the centre of the 100x100km domain. These data
are part of the $1km^{2}$ 1995 National Atmospheric Emissions
Inventory for 1995 \cite{Salway}. Other emission sources -
surrounding towns, large roads and point sources - were ignored as only
the effects of Edinburgh emissions are being examined. All areas
outside the city were assumed to have a low, background $NO_{x}$ emission.

Traffic count data from Edinburgh were used as a proxy for the
diurnal variations in $NO_{x}$ emissions. Approximately 85$\%$ of all $NO_{x}$ emissions were
estimated to come from road traffic in city centres
\cite{Lindqvist}, the rest was
assumed to be emitted by background sources and were treated as
constant throughout the 24 hours. The traffic count data used were from a detailed
city-centre traffic analysis, held over 16 hours (0600-2200hrs) at 40 sites
in June 1997, and annual average traffic
counts from seven main arterial routes into and out of the city, also for 1997.
As both the
city centre and surrounding arterial roads had similar diurnal variation,
the average of them was used in the model for $NO_{x}$ emission variations
- as seen in the composite plot in Figure 1.

\begin{figure}
\begin{center}
\includegraphics[width=12cm]{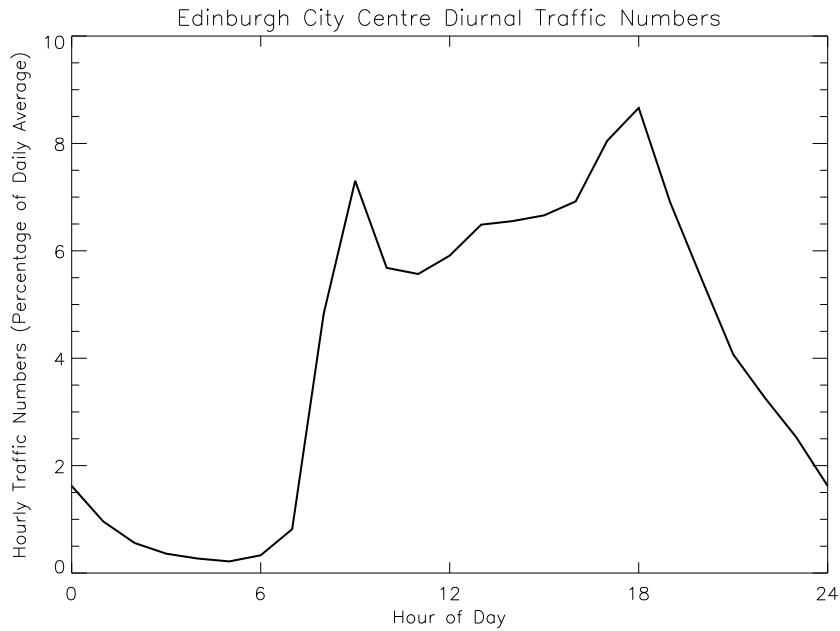}
\caption{Hourly traffic count data for Edinburgh city-centre,
1997 - Used as a proxy for $NO_{x}$ emissions}
\end{center}
\end{figure}

Rural background levels of $NO_{x}$ emissions ($\it{ie.}$ outside the 10x12km city
centre) were spatially constant and varied temporarily in the
same way as the city centre emissions. $NO_{x}$ emissions are put into the
lowest two layers of the model and a $NO:NO_{2}$ ratio of 3:1 is assumed.

\subsection{Wind Data}

The wind data used in these modelling studies were obtained from Turnhouse Airport, situated 8km west of
the city centre, and are averages for the period 1970-1991 \cite{CD}.
These are 10m winds and are used rather than the local geostrophic wind as they
are available in a detailed statistical form and yet are still 
representative of airflow across the region. The data include the speed 
and frequency (without
calm conditions) of the wind from different directions for midday and
midnight in June and
December (see Tables \ref{windsJune} and \ref{windsDec}).

\renewcommand{\baselinestretch}{1.0}
\small\normalsize

\begin{table}[h]
\begin{center}
\begin{tabular}{ccccc}
\hline
{DIRECTION} & \multicolumn{2}{c}{0000hrs} & 
\multicolumn{2}{c}{1200hrs}\\
     \cline{2-3} \cline{4-5}
 &Frequency ($\%$) & Speed (m s$^{-1}$) & Frequency($\%$) & Speed (m s$^{-1}$)\\
\hline
000$^{\circ}-044^{\circ}$ & 6.7 &2.3&9.4&3.7\\
$045^{\circ}-089^{\circ}$ & 19.1&3.5&28.0&4.9\\ 
$090^{\circ}-134^{\circ}$ & 9.6 &2.7&7.2&3.6\\
$135^{\circ}-179^{\circ}$ & 2.5 &2.0&2.1&2.5\\
$180^{\circ}-224^{\circ}$ & 6.0 &3.0&6.6&5.1\\
$225^{\circ}-269^{\circ}$ & 39.4&3.8&20.7&6.0\\
$270^{\circ}-314^{\circ}$ & 13.4&2.6&21.7&5.2\\
$315^{\circ}-359^{\circ}$ & 1.7 &1.7&4.4&3.3\\
\hline
 &100.0 & 3.0&100.0 & 5.0\\
\hline
\end{tabular}
\end{center}
\caption{Edinburgh wind-rose data (Turnhouse, 1971-1991) June 0000 and 1200hrs}
\label{windsJune}
\end{table}

\begin{table}[h]
\begin{center}
\begin{tabular}{ccccc}
\hline
{DIRECTION} & \multicolumn{2}{c}{0000hrs} &
\multicolumn{2}{c}{1200hrs}\\
     \cline{2-3} \cline{4-5}
 &Frequency ($\%$) & Speed (m s$^{-1}$) & Frequency($\%$) & Speed
(m s$^{-1}$)\\
\hline
000$^{\circ}-044^{\circ}$ & 3.8 &4.7&3.1&3.7\\
$045^{\circ}-089^{\circ}$ & 6.4&3.8&5.9&4.4\\
$090^{\circ}-134^{\circ}$ & 9.4 &5.0&10.3&5.7\\
$135^{\circ}-179^{\circ}$ & 2.3 &3.9&1.9&4.7\\
$180^{\circ}-224^{\circ}$ & 8.2 &7.4&10.0&5.3\\
$225^{\circ}-269^{\circ}$ & 49.1&6.0&43.5&6.2\\
$270^{\circ}-314^{\circ}$ & 18.4&3.9&21.3&4.4\\ 
$315^{\circ}-359^{\circ}$ & 2.6 &3.4&4.0&3.1\\
\hline
 & 100.0 &4.7 & 100.0 & 5.3\\
\hline
\end{tabular}   
\end{center}  
\caption{Edinburgh wind-rose data (Turnhouse, 1971-1991) December 0000 and
1200hrs}
\label{windsDec}
\end{table}

\renewcommand{\baselinestretch}{1.0}
\small\normalsize

\subsection{Observational Data}

Observations of ozone and $NO_{x}$ concentrations were used from three
monitoring sites - one urban: Princes Street, in
the city centre,
and two rural: Bush and Auchencorth Moss, to the south of the city. Details
of these sites are contained in Table \ref{sites}.
The inlet heights for monitoring at the three stations are different.
Bush and
Princes Street monitoring heights are at 4m and 5m respectively (in the
third level of the model) while Auchencorth is monitored at 3m (at the
top of the second level).
The output from all model runs are for level 3 in the column - except
for those that are looking at the variations in concentrations with
height as vertical profiles or sections.

\renewcommand{\baselinestretch}{1.0}
\small\normalsize

\begin{table}[h]
\begin{center}
\begin{tabular}{cccccc}
\hline
Station & Grid & Type & Site & Species & Measurement \\
        & Reference&  & Description & Analysed & Height \\
\hline
Princes Street & NT 254 738 & Urban & Urban Parkland, & $O_{3}$, $NO_{x}$, & 4m\\
 & & & 35m from major road. & $NO$ & \\
Bush Estate & NT 245 635 & Rural & Site surrounded by & $O_{3}$, $NO_{x}$, & 5m\\
 & & & open and forested land. & $NO$ & \\
Auchencorth & NT 221 562 & Rural & Moorland. Low local & $O_{3}$, $NO_{x}$, & 3m \\
 & & & agricultural activity. & $NO$ & \\
\hline
\end{tabular}   
\end{center}  
\caption{Location and description of the monitoring sites in and around Edinburgh}
\label{sites}
\end{table}

\renewcommand{\baselinestretch}{1.0}
\small\normalsize

In all three cases, the nitrogen oxide concentration, $[NO_{x}]$, and the
nitric oxide concentration, $[NO]$, are measured directly, and then the nitrogen
dioxide concentration is calculated by subtraction.

\section{One-Dimensional Trajectories Through a City}

\subsection{Method}

Straight-line trajectories, featuring rural and urban conditions, were
run over 100km to examine the effect on ozone 
concentration and the $O_{3}$/$NO$/$NO_{2}$ equilibrium
of different atmospheric conditions. Starting with constant
vertical concentrations at the upwind boundary, an initial fetch
of 45km
over background (rural) landscape allowed the $O_{3}$/$NO$/$NO_{2}$
system to reach a dynamic equilibrium and a vertical profile to form.
The column then passed over 10km of simulated urban area with a lower
deposition velocity from that over 
the rural area and (spatially constant) $NO_{x}$ emissions included. Finally, the 
trajectory continued
for another 45km downwind of the city, again over background rural
landscape, to allow the $O_{3}$/$NO$/$NO_{2}$ system to move towards
attaining a new 
balance. The amount to which the ozone concentration recovered was compared
with initial upwind concentrations. The output from the model is the
ozone concentration in the third vertical level (5m) of the column.

\subsection{Model Input data}

The one dimensional model runs were performed under four different
`extreme' atmospheric conditions - summer/winter and day/night - which
have large differences in their values of insolation, wind-speed and atmospheric
stability and thus show the range 
of ozone concentrations in seasonal and diurnal cycles.
As many factors as possible were
held constant between the different conditions so as to show which of the
many input parameters had the largest influence on the ozone results. 
Only the wind-speed, temperature (which has a negligible effect on ozone
concentrations with small variations) and solar insolation values
used were
different between daytime and nighttime runs in the same season. The `summer' and
`winter' and `day' and `night' conditions used in the  
modelling studies were based on four sets of meteorological parameters
outlined in Table \ref{1d-table}.

\renewcommand{\baselinestretch}{1.0}
\small\normalsize
 
\begin{table}[h]
\begin{center}
\begin{tabular}{ccccc}
\hline
 & Summer Day & Summer Night & Winter Day & Winter Night\\
\hline
  Date            &  15/6/97       & 15/6/97      & 15/12/97    &15/12/97\\
  Time            &  12:00hrs      & 00:00hrs     & 12:00hrs     &00:00hrs\\
  Wind-Speed      &  5.0m s$^{-1}$   & 3.0m s$^{-1}$  & 5.3m s$^{-1}$  &4.7m s$^{-1}$\\
  Temperature     & 15$^{\circ}$C & 10$^{\circ}$C & 5$^{\circ}$C &0$^{\circ}$C\\
  Cloud Cover     &  4oktas        & 4oktas       & 4oktas       &4oktas\\
  $NO/NO_{2}$ Ratio &  75:25           & 75:25          & 75:25    & 75:25\\
Initial [$NO$]&1.0$\mu$g m$^{-3}$&1.0$\mu$g m$^{-3}$&1.0$\mu$g m$^{-3}$&1.0$\mu$g m$^{-3}$\\
Initial [$NO_{2}$]&5.5$\mu$g m$^{-3}$&5.5$\mu$g m$^{-3}$&9.5$\mu$g m$^{-3}$&9.5$\mu$g m$^{-3}$\\
Initial [$O_{3}$]&90.0$\mu$g m$^{-3}$&90.0$\mu$g m$^{-3}$&75.0$\mu$g m$^{-3}$&75.0$\mu$g m$^{-3}$\\
\hline
\end{tabular}
\end{center}
\caption{Input parameters for the model for the simulated seasonal and 
diurnal trajectories} 
\label{1d-table} 
\end{table} 

\renewcommand{\baselinestretch}{1.0}
\small\normalsize

In each case, the initial concentration values was chosen so that, when run over 
background country, the model concentrations were close to the observed
average concentrations for those conditions at Auchencorth Moss - the
remote rural site.

The wind-speeds used were those
shown in the bottom rows in Tables \ref{windsJune} and \ref{windsDec},
$\it{ie.}$ the mean of the midday and midnight values so as to make the
daytime and nighttime model runs comparable.

The deposition velocity of ozone and $NO_{2}$ ($NO$ is not deposited at
a significant rate) for
each season was calculated by multiplying a seasonal average by a variable
factor.
At nighttime this factor had a value of 0.4 while in the daytime
it varied according to the zenith angle of the sun with a greater
amplitude in summer than winter.
The values of the deposition velocity used in the model are contained in
Table \ref{deposition}  \cite{Brook}.

\renewcommand{\baselinestretch}{1.0}
\small\normalsize

\begin{table}[h]
\begin{center}
\begin{tabular}{ccccc}
\hline
 & Summer Day & Summer Night & Winter Day & Winter Night\\
\hline
$V_{D}$ $O_{3}$ (Rural) &12.0mm $s^{-1}$&6.0mm $s^{-1}$&4.8mm $s^{-1}$&2.4mm $s^{-1}$\\
$V_{D}$ $O_{3}$ (Urban) &6.0mm $s^{-1}$&3.0mm $s^{-1}$&2.4mm $s^{-1}$&1.2mm $s^{-1}$\\
$V_{D}$ $NO_{2}$(Rural) &3.0mm $s^{-1}$&1.5mm $s^{-1}$&1.2mm $s^{-1}$&0.6mm $s^{-1}$\\
$V_{D}$ $NO_{2}$(Urban) &1.5mm $s^{-1}$&0.75mm $s^{-1}$&0.6mm $s^{-1}$&0.3mm $s^{-1}$\\\hline
\end{tabular}
\end{center}  
\caption{Annual and diurnal variations in the ozone and $NO_{x}$ deposition
velocities to rural and urban areas}
\label{deposition}
\end{table}

\renewcommand{\baselinestretch}{1.0}
\small\normalsize

\subsection{Results From One-Dimensional Trajectories}

Figure 2 shows the concentrations for summer day
and summer night trajectories - midday and midnight of June 15th respectively.

\begin{figure}
\begin{center}
\includegraphics[width=12cm]{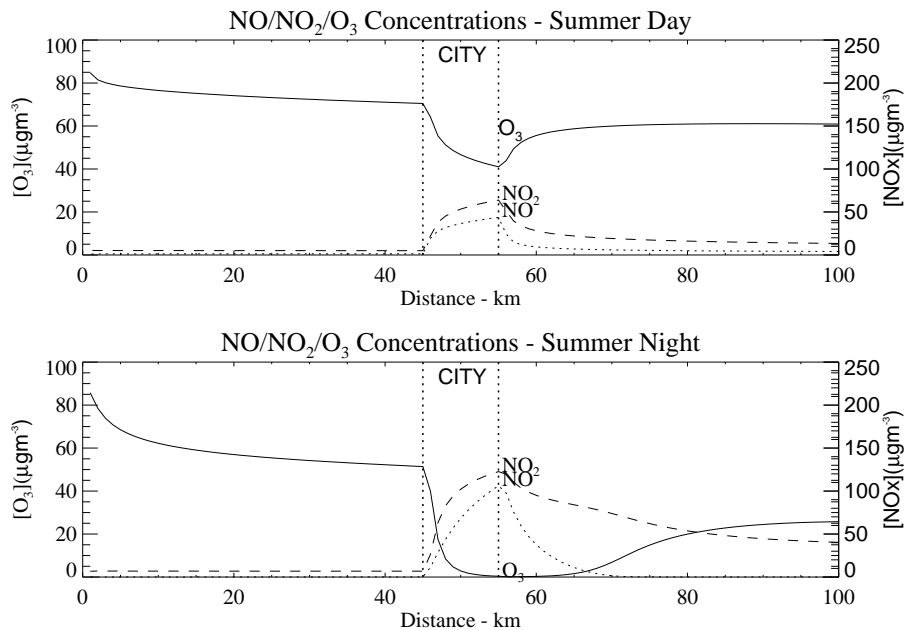}
\caption{Ground-level $O_{3}$, $NO$ and $NO_{x}$ concentrations
for a 100km straight-line trajectory featuring a 10km diameter city with
constant $NO_{x}$ emissions - summer day (top) and summer night (bottom).}
\end{center}
\end{figure}

At midday in summer, the concentration of ozone drops from an 
upwind value of 70$\mu$g m$^{-3}$ to a
minimum of 42$\mu$g m$^{-3}$ within the city boundaries. However,
owing to the rapid vertical mixing, once downwind of the city
the ozone concentration quickly recovers again towards 60$\mu$g m$^{-3}$ -
the maximum reached before deposition and chemistry start to reduce the
ozone again. This is an overall loss of 10$\mu$g m$^{-3}$ as the
air column crossed the city or about $15\%$ of the upwind concentration. Due
to the large amount of $NO_{x}$ emitted -
the new dynamic equilibrium of the $NO/NO_{2}/O_{3}$ system has larger 
concentrations of both $NO$ and $NO_{2}$.
At night the boundary layer is shallow
and stable which causes both $NO$ and $NO_{2}$
to accumulate, exceeding 100$\mu$g m$^{-3}$ by the downwind edge of
the city. The ozone within the city is completely destroyed and 
the $NO$ present close to the surface keeps the ozone concentration at
zero for almost 10km
downwind of the city. As there is no photolysis at night and thus no
reforming of $NO$, once the $NO$ emitted by the city has been
used up, the ozone concentration increases again due to the mixing down
of ozone-rich air from aloft. The stability of the lower atmosphere
ensures that this increase is relatively slow: by 30km downwind the
ozone concentration has reached
25$\mu$g m$^{-3}$, approximately $50\%$ of the value on the upwind edge of
the city.

Figure 3 shows two plots equivalent to Figure 2,
but for the mid-December instead of June. 

\begin{figure}
\begin{center}
\includegraphics[width=12cm]{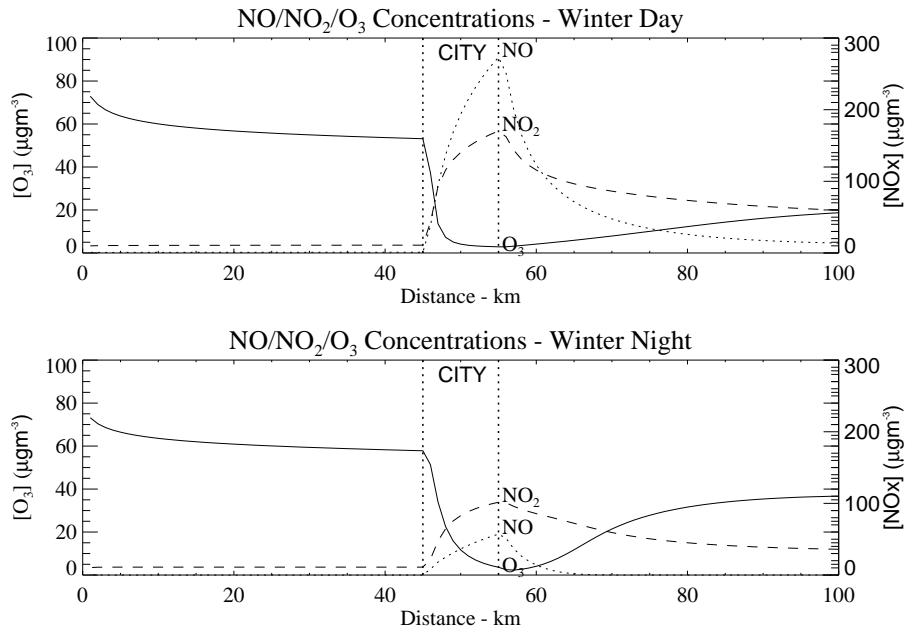}
\caption{Ground-level $O_{3}$, $NO$ and $NO_{x}$ concentrations
for a 100km straight-line trajectory featuring a 10km diameter city with
constant $NO_{x}$ emissions - winter day (top) and winter night
(bottom).}
\end{center}
\end{figure}

From Tables
\ref{windsJune}
and \ref{windsDec} it can be seen that the average
wind-speed is greater in winter than in summer, meaning
mechanical turbulence is greater. In December the incident
radiation is reduced, leading to less convective turbulence (and
hence less vertical mixing), and lower rates of photolysis. There are
also lower reaction rates and deposition velocities, but the emission
rates of $NO_{x}$ from the city are the same. It can be seen that in the
daytime, the greater
wind-speed doesn't overcome the lower convective turbulence levels and
leads to very high $NO$ and $NO_{2}$
concentrations in the city and hence less ozone
present than in June. Almost all of the 55$\mu$g m$^{-3}$
of ozone upwind of the city is destroyed while  $NO$ concentrations peak at
almost 300$\mu$g m$^{-3}$. The reduced vertical mixing due to greater
stability  of the boundary layer in the winter can be seen by
the time it takes for the $O_{3}$/$NO$/$NO_{2}$ system to
return towards a new dynamic equilibrium. At 45km downwind, both ozone and
$NO_{2}$ are still increasing as the destruction of all the $NO$ in the
lower layers of the column is slow. The increased wind-speed, and thus
mechanical turbulence, at midnight in December over the June values, together
with lower emissions, keeps the ozone
concentration slightly above zero and leads to a more rapid and complete
downwind recovery. As the $NO$
concentration decreases, the ozone concentration reaches 35$\mu$g m$^{-3}$
by 45km downwind - a $60\%$ recovery on the upwind concentration.

Figure 4 shows the vertical profiles of ozone concentration
over the lowest 150m of the boundary layer (first 10 levels in the model), taken at
three positions along the 100km trajectory. 

\begin{figure}
\begin{center}
\includegraphics[width=12cm]{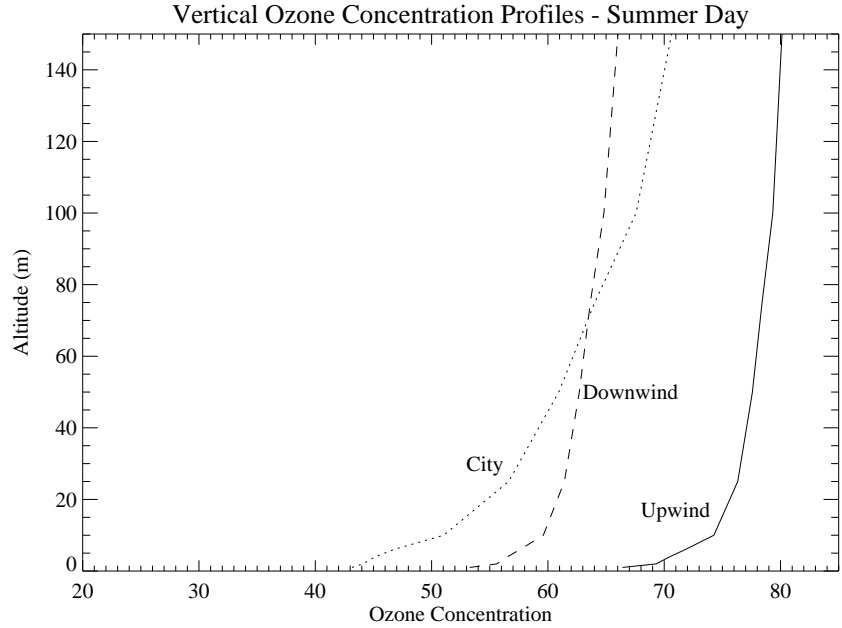}
\caption{Vertical profiles of ozone for summer daytime taken
upwind, within and downwind of a modelled city.}
\end{center}
\end{figure}

The profile labelled
`Upwind' is taken at the upwind boundary of the city $\it{ie.}$ after 45km of
rural emissions and deposition. The `City' profile was taken in the very middle of
the city (50km into the trajectory). It can be seen that in the middle
of the city ozone has been depleted from all levels of the column, but
especially from the lowest layers where the concentration is
$\>$20$\mu$g m$^{-3}$ less than in the upwind profile. The `Downwind' profile was
from 10km
downwind of the city. This profile shows a smaller
vertical 
concentration gradient throughout the column than in the city, indicating that the ozone concentration has
begun to recover near the foot of the column with the reduction of emissions,
but that the ozone depletion has
spread vertically upwards to deplete the higher levels. The profile is
almost parallel to the upwind profile with the depletion being around
12-15$\mu$g m$^{-3}$ throughout the column.

Figure 5 shows how the ozone concentrations under
different stability conditions depends on wind-speed. 

\begin{figure}
\begin{center}
\includegraphics[width=12cm]{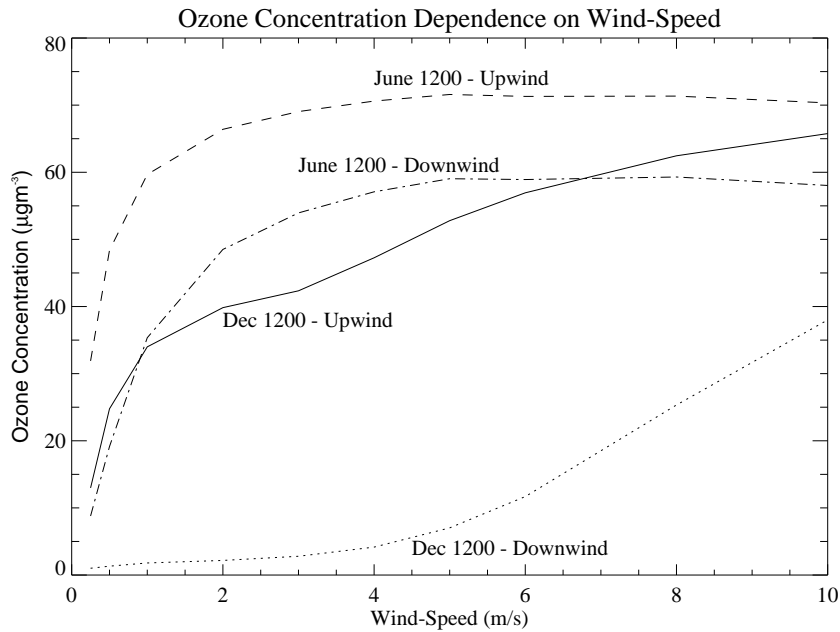}
\caption{Ozone concentrations (level 3) for summer and winter
midday trajectories taken upwind and downwind of the city under a range
of different wind speeds.}
\end{center}
\end{figure}

The ozone
concentrations were measured in the third level of the model at 10km upwind
(35km into
the trajectory) and 10km downwind (65km) under a range of wind-speeds varying
from 0.25 - 10m $s^{-1}$.
For the summer, both upwind and downwind, this diagram shows the expected
pattern - fairly
steady concentrations (independent of wind-speed) until the wind drops
below 2m $s^{-1}$ at which the concentration drops quickly
\cite{PORG97}. In
December, however, the model shows the concentration being much more dependent
on wind-speed over a larger range $\it{ie.}$ mechanical turbulence is more
important in winter and so the whole mixing is dependent on wind-speed
and hence so is the concentration. 

\section{Two-Dimensional Maps of Ozone Concentrations Within and Around Edinburgh}

\subsection{Method}

In order to simulate the two-dimensional field of ozone concentrations around
Edinburgh, the one-dimensional
trajectory version of the model has been modified to run over
a 100x100km grid, with Edinburgh located at the centre.
This has been achieved by combining the patterns from a series of
straight-line trajectories, from a variety of angles, so that
a picture is drawn up of the ozone concentration averaged over all directions.
The directions of the trajectories are separated from each other by 15$^{\circ}$,
giving 24 directions, each having its own frequency weighting and 
wind-speed, taken from the wind-rose data from section 2.4. 
The model is run at a 1x1km grid-scale. The ozone
concentration recorded at the end of each time-step is added to the
total of the grid-square currently occupied. When the whole domain has
been covered from each of the 24 angles used, the average
concentration of each grid-square is calculated by dividing the sum of
the concentrations for that square by the number of time-steps finishing
within it.

The initial $NO_{x}$ and $O_{3}$ concentration profiles used at the start of
each trajectory are calculated from the 100km summer/winter trajectories
from section 3. The concentrations at 20km of the trajectory over
rural land (with no emissions) were used. This allowed the chemistry to
reach a steady-state and vertical concentration profiles
that are appropriate for the atmospheric conditions to be created.

\subsection{Annual Mean Ozone Plot}

The annual mean ozone concentration map shows the effect to which
titration of ozone by nitric oxide has a pronounced effect throughout
the year. It was generated by averaging each
pixel output from twelve different model runs: midday and midnight for
October, December and February (using the `winter' conditions - $\it{ie.}$
wind-rose data and initial concentrations) and April,
June and August (using `summer' conditions) and
can be seen in Figure 6.

\begin{figure}
\begin{center}
\includegraphics[width=12cm]{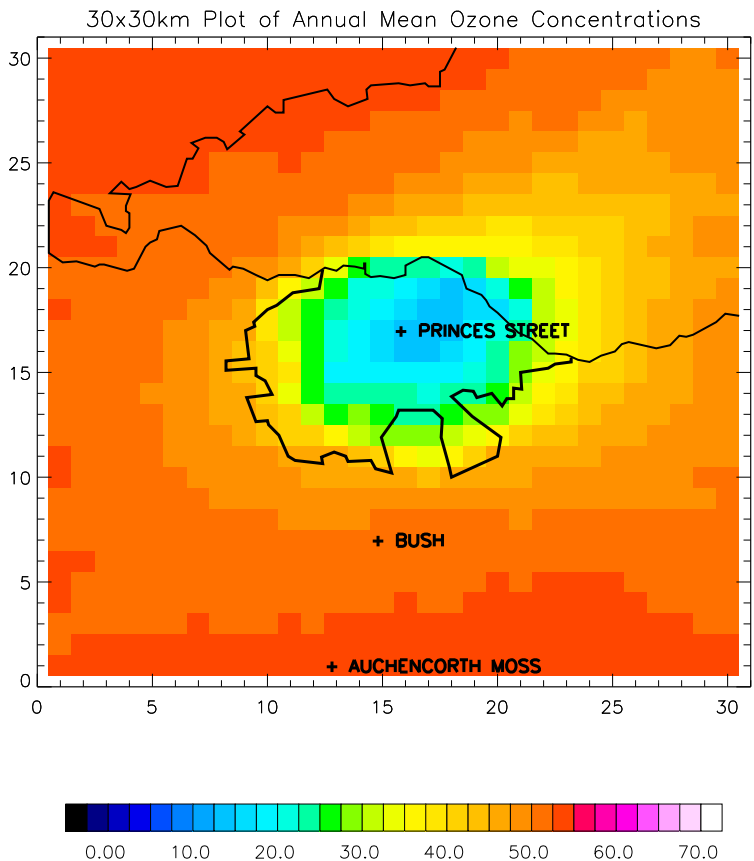}
\caption{30 x 30 km annual mean simulated ozone concentration for Edinburgh.}
\end{center}
\end{figure}

The output for the annual average ozone concentration at Bush and
Auchencorth are in excellent agreement with the annual average
observed
ozone concentration for the two sites in 1995-97 - see Table
\ref{Ann_Av_table}. It should be remembered, however, that Auchencorth
values were used to determine the upwind boundary conditions.

\renewcommand{\baselinestretch}{1.0}
\small\normalsize
 
\begin{table}[h]
\begin{center}
\begin{tabular}{ccc}
\hline
 SITE & MODELLED & OBSERVED \\
\hline
 Princes Street & 16$\mu$g m$^{-3}$ & 31$\mu$g m$^{-3}$\\
 Bush & 54$\mu$g m$^{-3}$& 53$\mu$g m$^{-3}$\\
 Auchencorth Moss & 55$\mu$g m$^{-3}$ & 54$\mu$g m$^{-3}$\\
\hline
\end{tabular}
\end{center}
\caption{Annual average ozone concentrations at the three monitoring
sites - both modelled and observed results}
\label{Ann_Av_table}
\end{table}

\renewcommand{\baselinestretch}{1.0}
\small\normalsize

However, at Princes Street the agreement is not nearly so good with the
modelled value only about half the observed average; but the model
value is the average concentration over the whole 1x1km
grid-square that the monitoring station is in. In reality the concentration
varies greatly across
the grid square and the observations may be taken at a site where the
concentration is not representative of the whole area. This is in
contrast to the Auchencorth site, where the uniformity of the landscape and the
lack of local sources makes it representative of a large area. 
It might though be expected that the observations, particularly at
Princes Street, would be smaller than the modelled results as they are
taken within a few metres of the road itself. The position of the
monitoring site though is slightly unusual in that it is
positioned within a garden that is at a lower level than the road.
The model treats the entire domain as it it were flat and is unable 
to recreate the effects of such topographic detail.
However, with the height of the monitoring inlet being 4m above the 
surrounding ground, it is almost
on the same level as the adjacent road and thus should still be sampling
air that is depleted in ozone. It appears that the model is 
under-estimating ozone concentrations within urban areas.

The spatial extent of ozone depletion greater than 5$\mu$g m$^{-3}$ from
the background (Auchencorth) value of 55$\mu$g m$^{-3}$ is
fairly limited on an
annual-average basis. The 50$\mu$g m$^{-3}$ isopleth extends only a few
kilometres outside the city boundary - reaching a maximum distance of
approximately 10km downwind of the dominant wind direction (see
Tables \ref{windsJune} and \ref{windsDec}), to the north-east. Areas
with concentrations less than
45$\mu$g m$^{-3}$ (a 20$\%$ depletion) are mainly confined within the city boundary and, due
to the prevailing wind from the southwest, out over the Firth of Forth
to the north-east.
Only an area of approximately 8x8km, just offset from the city centre is depleted
by more than 50\%.

\section{Modelling Vertical and Temporal Variations in Edinburgh's Ozone Concentrations}

\subsection{Vertical Section of Ozone Concentrations Through Edinburgh}

Figure 7 is the 30x30km plot of ozone concentrations for
midday in December.

\begin{figure}
\begin{center}
\includegraphics[width=10cm]{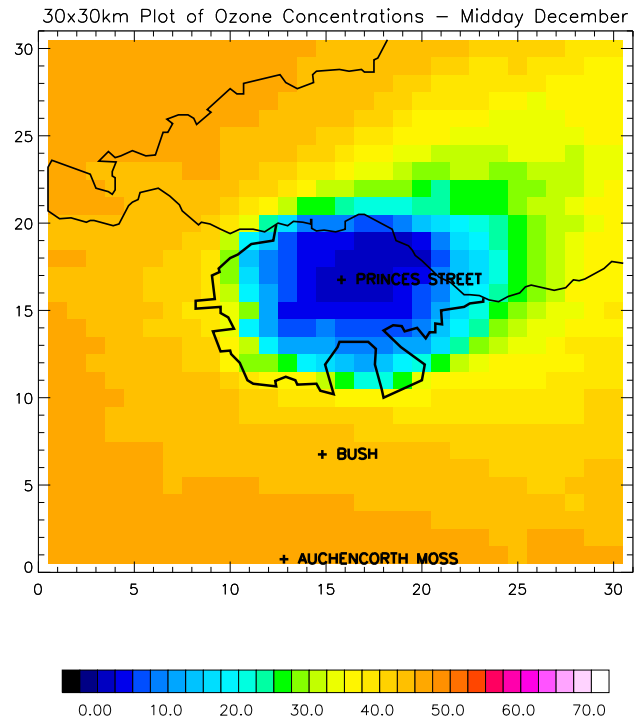}
\caption{30 x 30 km simulated ozone concentration for Edinburgh -
December, 1200h.}
\end{center}
\end{figure}                                                            

The city centre
ozone concentration is less than 5$\mu$g m$^{-3}$ with the Bush and
Auchencorth concentrations being 46 and 48$\mu$g m$^{-3}$ respectively.
Depletion greater than 20$\%$ ($\it{ie.}$ less than 35$\mu$g m$^{-3}$) is 
again mainly confined to the city and areas
to the north and east, out over the Firth of Forth.

Figure 8 is a north-south vertical
section (0-150m) through the 1200hrs December ozone concentration field
- passing through Princes Street. Above 100m there is little detail of the city
below that can be seen at all (except for some jagged peaks that are artefacts
of the plotting program used). It can be seen that the section is skewed with
lower concentration spread out towards the north - this can also be seen in
the two-dimensional plot (Figure 7). Concentrations of less
than 25$\mu$g m$^{-3}$, $\it{ie.}$ 50\% depletion,
are restricted to below 20m.
At nighttime, this would be even lower - just a few metres above the ground.

\subsection{Diurnal Variation of Ozone Concentrations}

Diurnal plots of ozone concentrations at Princes Street and Bush for
December have been made by calculating the hourly ozone concentrations
for 24 hours. These are plotted in Figure 9 and are compared
with observational data obtained from the two monitoring stations
(hourly averages) from 3 years (1995-97). Any day with missing data for
any of the species were ignored; however overall
data capture was greater than $90\%$.

As can be seen from the diurnal variation plot, the values at Bush
generated by the model are closer to the observed average values (as
expected due to the definition of the upwind perimeter concentrations)
than those at Princes Street but the pattern of the diurnal variation
produced by the model at Princes Street
more closely resembles the observations, despite a systematic
under-estimation of 10-20$\mu$g m$^{-3}$.

At Princes Street, the only feature of the modelled variation that
doesn't correspond well with the observed values is in the middle of the day
when there is a peak in the concentration from the minima during the
dawn and dusk rush hours. This is due to the increased mixing in
the day in the model from the increased insolation. The reason that the
modelled values are on average 10$\mu$g m$^{-3}$ less than those observed
may be due to one or more of several reasons:

\begin{itemize}

\item The emission used in the model may not be appropriate for the locality.

\item Vertical mixing may be underestimated. No account is taken of the heat
island effect.

\item It is likely that the complex mixing processes taking place within the
street canyons are not adequately represented.

\item Mixing depends on stability, which is determined only by external
conditions (cloud cover and wind speed) at the time. There is no
"memory" of preceeding conditions.

\end{itemize}

The observed values at Bush are lower than those produced by the model
and show that this site is also influenced by the
Edinburgh rush hour. There are two roads within 500m of the site
that carry substantial traffic into and out of the city in the morning and
evening and the corresponding dips in the ozone concentration at 0900 and
1700hours can be attributed to this. There appears to be a discontinuity in the
modelled data in the very middle of the day, where the concentration
dips instead of rising to a small peak at midday as happens at Princes Street.
This may be due to the ozone deposition velocity during the
day over rural areas reaching such a value that depletion of ozone at
the surface exceeds the mixing down of ozone rich air from aloft.
The small deposition velocity and greater mechanical turbulence over the
city ensure that this does not occur at the Prices Street site.

\section{Conclusions}

The Lagrangian column model has been used in this study to produce a series
of one- and two-dimensional maps of surface ozone concentrations 
under a variety of meteorological conditions.

One-dimensional straight-line trajectories under average summertime
boundary-layer conditions show a depletion of surface ozone over an
urban area of $40\%$ in
daytime with a downwind recovery to approximately $80\%$ of upwind rural
values at 20km and $100\%$ depletion with an eventual recovery to
$50\%$ downwind at night.
Similar trajectories run in winter, with lower insolation levels and
greater mean wind-speeds show $90\%$ of the surface ozone is depleted in
both daytime and nighttime conditions. At nighttime, the downwind
recovery to $60\%$ of the upwind concentration is faster than the
daytime recovery.

Vertical profiles upwind, within and downwind of the city show how the
concentration gradient in the lowest 150m increases as the column passes
over the city and then decreases again downwind as the
loss of ozone is evenly spread through the layers. The upwind and
downwind profiles have very similar shapes - at midday in June the
difference between the two profiles is approximately 12$\mu$g m$^{-3}$
throughout the profile.
A vertical section through the simulated city for December shows that
$50\%$ depletion is restricted to the lowest 20m of the vertical column
and that at 100m altitude, the effects of the city on ozone
concentrations are barely discernible.

Two-dimensional maps of annual mean surface ozone concentrations show that the
depletion of ozone by a city is restricted to very close to the urban
area. A depletion of $20\%$ of the background ozone in a simulated
Edinburgh only
extends beyond the city boundaries downwind of the prevailing wind
direction while $50\%$ depletion is completely confined within the
city boundaries.

At this stage we can revisit the omission of ozone generation from the
model formulation. Typical gradients of ozone concentration around the
city boundary are 20-30$\mu$g m$^{-3}$ per 5km (see Fig. 6).
Observations of ozone generation rates are typically at least an order of
magnitude less than this (e.g. \cite{Weston}), so that patterns of
concentration would not be significantly affected by its inclusion.

The diurnal cycle of ozone concentrations at urban and rural sites has been
compared with monitoring site data. The model recreates the shape of the
diurnal cycle well - but consistently underestimates the city-centre
surface concentrations. This could be due to underestimating atmospheric
turbulence,
overestimating emissions or the fact that the single point of the monitoring
station does not accurately represent the full 1x1km grid-square. The model 
also overestimates the
concentrations at the rural site, which is more dependent on the
emissions from local roads than used in the model.

\section{Acknowledgements}

The authors acknowledge Dr Rod Singles and Dr Helen ApSimon et al. for
the use of the TERN model and their expert advice. Ms Mhairi Coyle, Dr Chris
Flechard (C.E.H., Edinburgh), Dr Justin Goodwin (NETCen) and Edinburgh City
Council Roads Department are thanked for supplying data. J. Nicholson was
in receipt of a Natural Environment Research Council
studentship.

\newpage
\bibliography{main}

\end{document}